\documentclass[conference]{IEEEtran}
\IEEEoverridecommandlockouts
\usepackage{cite}
\usepackage{amsmath,amssymb,amsfonts}
\usepackage{algorithmic}
\usepackage{graphicx}
\usepackage{textcomp}
\usepackage{xcolor}
\usepackage{minted}
\usepackage{listings}

\newcommand{\sect}[1]{Sec.\ref{sec:#1}}

\def\BibTeX{{\rm B\kern-.05em{\sc i\kern-.025em b}\kern-.08em
    T\kern-.1667em\lower.7ex\hbox{E}\kern-.125emX}}
\begin{document}

\title{High-Order Spectral Element Methods for Wave Propagation on ARM Multicore CPU with SME: Optimizations and Implications
}

\author{\IEEEauthorblockN{Yinuo Wang}
\IEEEauthorblockA{\textit{Tsinghua University} \\
Beijing, China\\
wyn22@mails.tsinghua.edu.cn}
\and
\IEEEauthorblockN{Lin Gan}
\IEEEauthorblockA{\textit{Tsinghua University} \\
\textit{Hetao Institute of Mathematics and Interdisciplinary Sciences, Shenzhen}\\
Beijing, China \\
lingan@tsinghua.edu.cn}
\and
\IEEEauthorblockN{Tianqi Mao}
\IEEEauthorblockA{\textit{Tsinghua University} \\
Beijing, China \\
mtq24@mails.tsinghua.edu.cn}
\and
\IEEEauthorblockN{Wubing Wan}
\IEEEauthorblockA{\textit{Tsinghua University} \\
Beijing, China \\
wwb21@mails.tsinghua.edu.cn}
\and
\IEEEauthorblockN{Zekun Yin}
\IEEEauthorblockA{\textit{Shandong University} \\
Jinan, Shandong \\
Zekun.yin@sdu.edu.cn}
\and
\IEEEauthorblockN{Wenqiang Wang}
\IEEEauthorblockA{\textit{National Supercomputing Center in Shenzhen} \\
Shenzhen, China \\
wangwq2018@mail.sustech.edu.cn}
\and
\IEEEauthorblockN{Wei Xue}
\IEEEauthorblockA{\textit{Tsinghua University} \\
Beijing, China \\
xuewei@tsinghua.edu.cn}
\and
\IEEEauthorblockN{Guangwen Yang}
\IEEEauthorblockA{\textit{Tsinghua University} \\
Beijing, China \\
ygw@tsinghua.edu.cn}
}

\maketitle

\begin{abstract}
Wave propagation based on the spectral element method (SEM) is a representative HPC workload, but existing SEM implementations are not well matched to emerging ARM multicore CPUs with Scalable Matrix Extension (SME). We present an SME-enabled optimization of \textsc{SPECFEM3D} on the emerging LX2 processor that combines an SME-aware batched small-matrix kernel for SEM tensor-product operators, a memory-aware hybrid MPI+OpenMP execution scheme for limited-HBM systems, and a dispersion-based iso-accuracy study of the $(h,p)$ tradeoff. At fixed polynomial order, the optimized implementation improves full-application performance by 4--6$\times$ over the original code and delivers clear gains over optimized non-SME CPU baselines. Beyond these implementation-level gains, our results suggest that SME shifts the performance-favorable operating point toward higher polynomial orders along the dispersion-based iso-accuracy frontier, further reducing time-to-solution and working-set size. These results indicate that SME affects not only kernel efficiency, but also the practical discretization tradeoff for SEM on modern ARM multicore platforms.
\end{abstract}


\section{Introduction}
Wave propagation plays a central role in earth modeling, earthquake hazard assessment~\cite{Zhang2014, Chen2018, Wan2023}, and subsurface resource exploration~\cite{Zhou2018, Li2017}. Because of both its scientific importance and its exceptional demands on computational power and parallelism, numerical wave-propagation simulation has long been a representative large-scale workload in high-performance computing. The pursuit of accurate and efficient modeling of wave propagation in complex media and geometries dates back to the early development of supercomputers and remains an active challenge today.

To meet the need for high accuracy and efficiency in the presence of complex media, irregular geometries, and possible multi-physics coupling, the spectral element method (SEM) has become one of the most promising numerical approaches~\cite{Komatitsch1999, Komatitsch2002, Canuto2007}. SEM combines high-order polynomial basis functions with tensor-product Gauss-Legendre-Lobatto interpolation on polytope elements, achieving high numerical accuracy together with strong computational efficiency. Compared with classical low-order finite element methods, SEM provides superior accuracy per degree of freedom while retaining the geometric flexibility needed to handle complex domains, boundary conditions, and coupled physics.

From an HPC perspective, SEM offers several intrinsic advantages for large-scale simulation. By employing tensor-product nodal bases with Gauss-Legendre-Lobatto quadrature, it yields a diagonal mass matrix, which eliminates the need for costly mass-matrix inversion and enables efficient explicit time integration. In addition, because only shared nodes on partition boundaries require data exchange, communication is restricted to nearest neighbors, which helps preserve scalability in distributed-memory execution.

SEM also benefits from a matrix-free formulation~\cite{Germann2021, Swirydowicz2019}, which avoids low-efficiency sparse matrix operations and instead evaluates element contributions on the fly. This reduces memory traffic and better matches the arithmetic structure of high-order methods. As a result, SEM maps naturally to modern SIMD/SIMT architectures and can achieve higher hardware utilization than traditional sparse finite-element formulations.

These properties have made SEM the foundation of several influential high-performance simulation frameworks and applications, including \textsc{SPECFEM3D}~\cite{Komatitsch2011}, \textsc{MFEM}~\cite{Andrej2024}, and \textsc{SEM3D}~\cite{Touhami2022}, and have contributed to its strong reputation in large-scale scientific computing~\cite{Bell2017, Henneking2025}.

Historically, most numerical frameworks were developed under older hardware assumptions or optimized primarily for accelerator-centric environments such as GPU cluster. In contrast, HPC is now undergoing a renewed architectural shift toward large-scale ARM multicore CPU platforms equipped with advanced vector and matrix extensions. For example, China is advancing its national HPC infrastructure with the next-generation ARM-based exascale-class system Lingsheng at the National Supercomputing Center in Shenzhen.  Together with established ARM-based flagship systems such as Japan’s Fugaku~\cite{Sato2022} and its planned successor FugakuNEXT~\cite{Riken2025}, this trend indicates that ARM multicore CPUs will play an increasingly important role in future leading HPC systems.

In particular, the ARM Scalable Matrix Extension (SME) introduces matrix-oriented instructions based on outer-product computation and tile storage, providing dedicated hardware support for matrix multiplication~\cite{ARM_SME}. From a hardware perspective, SEM appears to be a natural target for SME, since its stiffness operator is dominated by batched small matrix multiplications arising from tensor-product differentiation and the weak-form operator application. However, this opportunity is not realized directly in conventional SEM configurations. For the polynomial orders commonly used in practice, the resulting matrix sizes are too small to utilize SME tiles efficiently, and therefore a naive SME implementation is often unable to deliver a substantial advantage over conventional SIMD-based kernels.

To address this challenge, we develop an SME-enabled batched small matrix multiplication kernel based on software pipelining and vector aggregation, which improves SME tile utilization and achieves higher performance than traditional SIMD-backed implementations. More importantly, once this kernel-level inefficiency is removed, the limiting factor is no longer only the microarchitectural mapping, but also the application-level discretization choice. For wave-propagation problems, different combinations of element size \(h\) and polynomial order \(p\) form an iso-accuracy frontier w.r.t. dispersion error. Since SME changes the efficiency of the dominant batched matrix multiplication kernels along this frontier, we observe that its introduction shifts the performance-favorable operating point toward the higher-\(p\) side. We validate this observation through dispersion-based iso-accuracy wave-propagation experiments in homogeneous media.In the tested equal-accuracy regime, higher-p discretizations paired with SME-enabled kernels reduce both time-to-solution and working-set size.

To support realistic wave-propagation simulations, another major challenge must be addressed: memory capacity. Beyond the distributed field variables, practical SEM applications such as \textsc{SPECFEM3D} typically require each rank to maintain substantial auxiliary data, including tomographic medium descriptions, surface geometry information, preprocessing metadata, and communication buffers. Under the traditional MPI-only execution model adopted by many existing SEM frameworks, these per-rank data structures are replicated across a large number of processes. On modern ARM multicore platforms with very high core counts, such replication places severe pressure even on the DDR memory subsystem. During the PDE solution stage, additional memory is consumed by halo buffers, duplicated boundary degrees of freedom, and process-local runtime metadata, further reducing the limited HBM capacity available on each NUMA domain.

We therefore develop a hybrid MPI+OpenMP parallelization scheme for SEM on multicore CPUs. MPI processes are mapped to NUMA domains, while thread-level parallelism is exploited within each domain to reduce per-process memory replication. To eliminate shared-memory race conditions during nodal updates, we apply graph coloring instead of relying on costly atomic operations. At the inter-process level, we introduce a dedicated communication thread to achieve fully asynchronous neighbor exchanges and overlap communication with computation. This hybrid execution scheme substantially reduces memory consumption and enables realistic wave-propagation simulations on high-performance multicore CPUs.

In this work, we select \textsc{SPECFEM3D} as a representative wave-propagation application and implement our proposed optimizations on LX2, the processor powering China’s new supercomputer \emph{Lingsheng}. Our optimized implementation achieves a full-application performance improvement of $4$--$6\times$ over the baseline, while enabling realistic wave-propagation workloads under the memory constraints of the target ARM multicore platform. Our main contributions are summarized as follows:

\begin{enumerate}
    \item We design an SME-aware batched small-matrix kernel for SEM tensor-product operators on ARM multicore CPUs. By combining software pipelining with layout-aware vector aggregation, the proposed kernel improves utilization of SME tiles for the practically relevant small operator sizes arising in SEM wave propagation problem.

    \item We show that SME changes the practical efficiency trend of SEM across polynomial orders, making higher-$p$ discretizations substantially more attractive on CPUs, and experimentally use a dispersion-based iso-accuracy study to show that SME shifts the hardware cost structure of the dominant tensor-product operators and thus motivates renewed consideration of higher-order SEM on modern ARM CPUs.
    
    \item We develop a memory-aware hybrid MPI+OpenMP execution scheme for SPECFEM3D on multicore NUMA systems. This design reduces per-process memory replication and supports realistic wave-propagation workloads under limited HBM capacity, using graph coloring to remove shared-memory update conflicts and a dedicated communication thread to overlap neighbor exchange with computation.

\end{enumerate}

\section{Background}
\subsection{LX2 Architecture}  \label{sec:kp920f}
\begin{figure}[h]
     \centering
     \includegraphics[width=\linewidth]{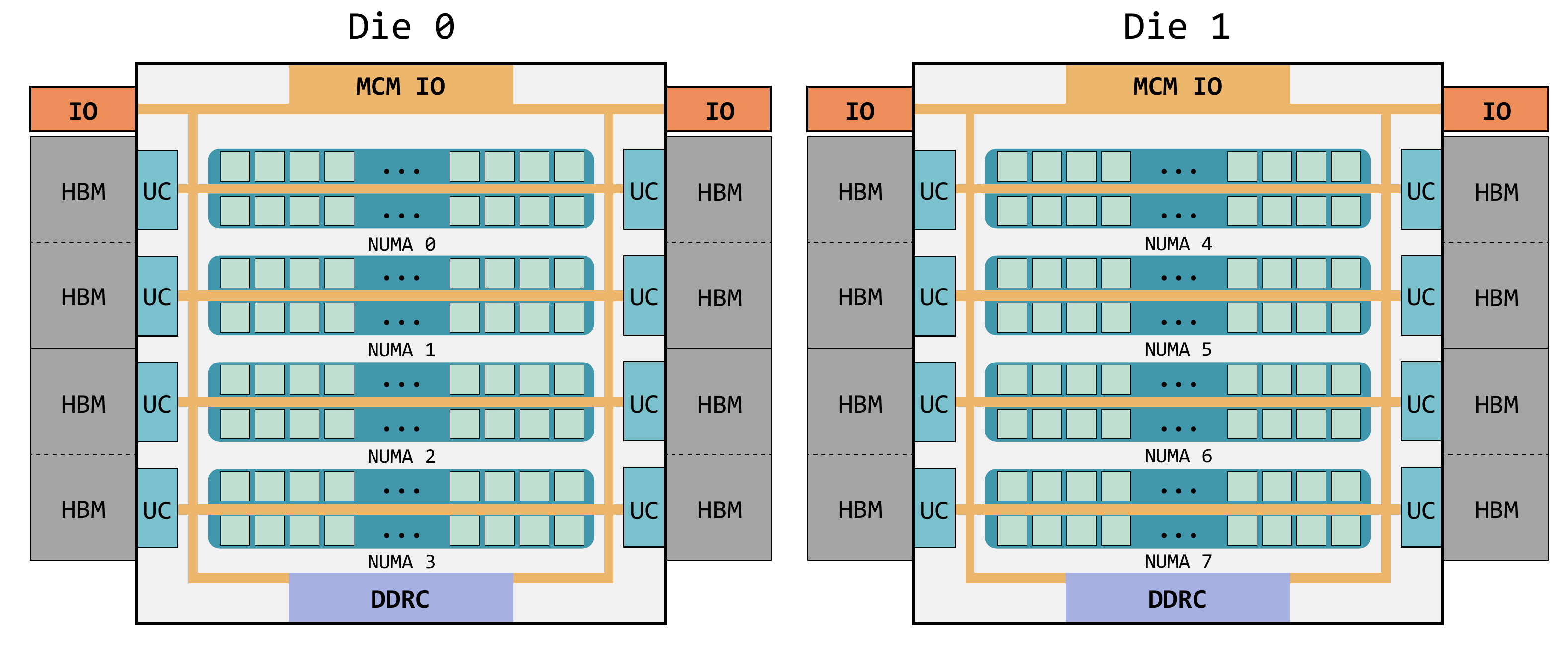}
     \caption{LX2 CPU Architecture}
     \label{fig:kp920f}
\end{figure}
The LX2 is a dual-die ARMv9-A processor featuring 304 physical cores operating at 1.6 GHz. Each die comprises four NUMA domains of 38 cores. Every core includes private L1/L2 caches and supports the Scalable Matrix Extension (SME) with specialized outer-product units.

The memory subsystem features 4 GB of High Bandwidth Memory (HBM) per NUMA node (500 GB/s), configurable as either a hardware-managed cache or a software-addressable flat memory. Each die also accesses 1 TB of shared DDR memory (120 GB/s). A 160-channel System Direct Memory Access (SDMA) engine facilitates asynchronous transfers between DDR and HBM, effectively overlapping memory traffic with computation.

For intensive matrix computations, the SME unit employs a vector outer-product model. It computes the outer product of two vectors loaded from SVE registers, accumulating the result into a $64 \times 64$-byte 2D architectural register known as the ZA tile. The unit maintains full IEEE floating-point compliance. Data moves flexibly between the ZA tile and SVE registers via horizontal or vertical slice instructions. By interleaving operations to hide instruction latency, SME delivers approximately $4\times$ the theoretical peak performance of standard SVE Multiply-Accumulate (MLA) instructions.

\subsection{Spectral Element Method} \label{sec:sem}
\begin{figure}[h]
     \centering
     \includegraphics[width=\linewidth]{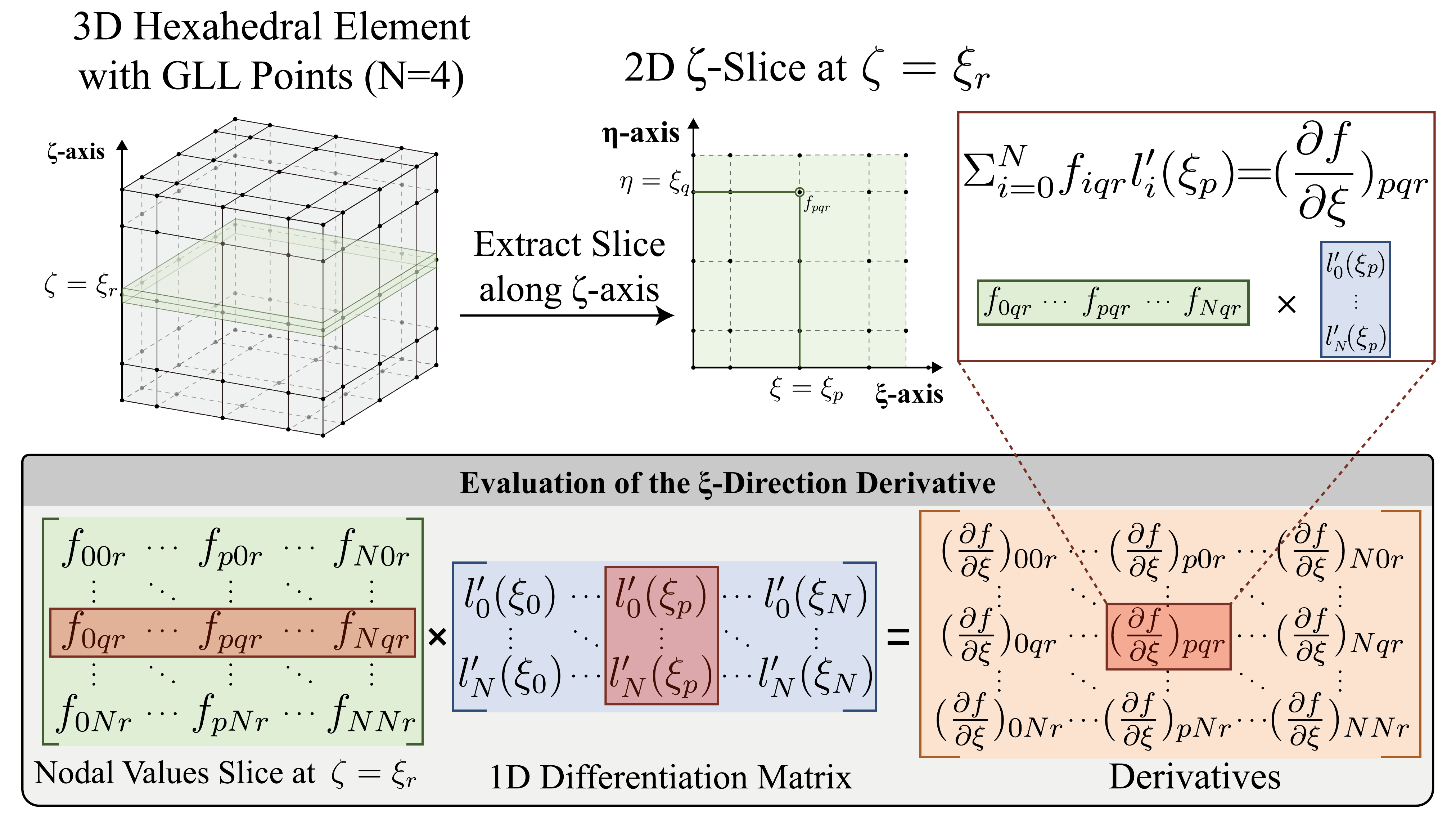}
     \caption{$\xi$-Derivative in Tensor-Product Element Basis}
     \label{fig:tp-derivatives}
\end{figure}

The spectral element method (SEM) is a class of high-order finite element methods. 
On a three-dimensional mesh it typically employs hexahedral elements. 
Within each element on the reference domain with coordinates $(\xi,\eta,\zeta)$, 
the interpolation (nodal) points are defined as the tensor product of one-dimensional 
Gauss-Legendre-Lobatto (GLL) points. 
If $\{\xi_i\}_{i=0}^N$ denotes the set of $(N+1)$ one-dimensional GLL nodes, 
then the coordinates of the three-dimensional interpolation nodes in reference space are
\[
  \boldsymbol{x}_{ijk} = \bigl(\xi_i,\ \xi_j,\ \xi_k\bigr),
\]
for $i,j,k = 0,\dots,N$.

The corresponding nodal basis functions within each element are given by the tensor product 
of one-dimensional Lagrange interpolation polynomials:
\begin{equation*}
    l_{ijk}(\xi,\eta,\zeta) = l_i(\xi)\, l_j(\eta)\, l_k(\zeta),
\end{equation*}
where $l_i(\xi)$ is the $i$-th one-dimensional Lagrange basis polynomial associated with the 
GLL nodes $\{\xi_0,\xi_1,\dots,\xi_N\}$ for a polynomial of degree $N$. 
It can be written explicitly as
\begin{equation*}
    l_i(\xi) 
    = \frac{(\xi - \xi_0)\cdots(\xi - \xi_{i-1})(\xi - \xi_{i+1})\cdots(\xi - \xi_N)}
           {(\xi_i - \xi_0)\cdots(\xi_i - \xi_{i-1})(\xi_i - \xi_{i+1})\cdots(\xi_i - \xi_N)}.
\end{equation*}
By construction, these basis functions satisfy the Kronecker delta property
\begin{equation*}
    l_i(\xi_j) = \delta_{ij}, \qquad i,j = 0,\dots,N,
\end{equation*}
which implies $l_{ijk}(\xi_{p},\eta_{q},\zeta_{r}) = \delta_{ip}\,\delta_{jq}\,\delta_{kr}$ 
for the three-dimensional tensor-product basis.

This tensor-product interpolation basis has a particularly convenient structure. 
For a function with nodal values 
$f_{ijk} = f(\xi_i,\xi_j,\xi_k)$, 
its value at an arbitrary point $(\xi,\eta,\zeta)$ in the reference element can be written as
\begin{equation*}
    f(\xi,\eta,\zeta) 
    = \sum_{i,j,k=0}^{N} f_{ijk}\, l_i(\xi)\, l_j(\eta)\, l_k(\zeta).
\end{equation*}
Taking derivatives in $\xi$-direction with respect to the reference coordinates yields
\begin{equation*}
    \frac{\partial f}{\partial \xi}(\xi,\eta,\zeta) 
    = \sum_{i,j,k=0}^{N} f_{ijk}\, l'_i(\xi)\, l_j(\eta)\, l_k(\zeta)
\end{equation*}
At the interpolation points $(\xi_p,\xi_q,\xi_r)$, the Kronecker-delta property 
$l_j(\xi_q)=\delta_{jq}$ and $l_k(\xi_r)=\delta_{kr}$ implies that, in each derivative, only nodes aligned 
with $\xi$-axis contribute to the result, while nodes with 
mismatched indices have no contribution. Consequently, evaluation of the $\xi$-direction derivative can be expressed as a sequence of small matrix-matrix multiplications of an $(N+1)\times(N+1)$ one-dimensional differentiation matrix with an 
$(N+1)\times(N+1)$ slice of nodal values taken along the $\xi$-direction, 
and similarly for $\eta$ and $\zeta$ (see Fig.~\ref{fig:tp-derivatives}).

For the representative application \textsc{SPECFEM3D} we consider, the governing equation is the elastic wave equation
\begin{equation*}
    \rho\, \partial_t^2 \mathbf{s}
    = \boldsymbol{\nabla} \cdot \mathbf{T} + \mathbf{f},
    \qquad
    \mathbf{T} = \mathbf{C} : \boldsymbol{\nabla}\mathbf{s},
\end{equation*}
where $\rho$ is the density of the solid, $\mathbf{s}$ is the displacement field, 
$\mathbf{T}$ is the stress tensor governed by Hooke's law, and $\mathbf{C}$ is the fourth-order stiffness tensor.
The corresponding weak form reads
\begin{equation*}
    \int_{\Omega} \rho\, \mathbf{w} \cdot \partial_t^2 \mathbf{s}\, \mathrm{d}^3\mathbf{x}
    = - \int_{\Omega} \boldsymbol{\nabla}\mathbf{w} : \mathbf{T}\, \mathrm{d}^3\mathbf{x}
      + \int_{\Omega} \mathbf{w} \cdot \mathbf{f}\, \mathrm{d}^3\mathbf{x},
\end{equation*}
for all test functions $\mathbf{w}$.

In the SEM discretization, both the trial and test functions are chosen as the tensor-product
Lagrange basis functions described above. The left-hand side gives rise to the \emph{mass matrix};
with inexact numerical integration at the Gauss-Lobatto-Legendre nodes, this matrix is diagonal to machine precision.
The main computational hot spot is the first term on the right-hand side, which corresponds to the \emph{stiffness matrix}.
SEM employs a matrix-free approach to evaluate the action of this stiffness operator.

The computation of the stress tensor $\mathbf{T}$ involves spatial derivatives of $\mathbf{s}$ in the three reference
directions, which, due to the tensor-product structure, can be expressed as three small batched matrix-matrix
multiplications. Applying the weak form, i.e., contracting with the gradients of the test functions,
leads to an additional three small batched matrix-matrix multiplications of the same structure.

\section{Related Work}

A diverse range of frameworks exists for spectral element methods (SEM) in wave propagation, including specialized packages such as \textsc{SEM3D}~\cite{Touhami2022} and \textsc{SPECFEM3D}~\cite{Komatitsch2011}, as well as more general high-order finite-element libraries such as \textsc{MFEM}~\cite{Andrej2024} and \textsc{deal.II}~\cite{Arndt2021}, which prioritize generality and performance portability. Among these, \textsc{SPECFEM3D} stands out as one of the most influential frameworks for wave propagation. This widely used, open-source seismic simulation tool, based on the spectral-element method, is designed for large-scale applications, ranging from local and regional models to global Earth simulations. Its significance in both the seismology and HPC communities is underscored by its receipt of the ACM Gordon Bell Award in 2003, its selection as a Gordon Bell finalist in 2008, and its distribution through the Computational Infrastructure for Geodynamics (CIG)~\cite{Bell2017}.

More broadly, high-order tensor-product FEM/SEM has garnered sustained attention in the HPC community due to its dense operator structure and matrix-free formulation, which align well with modern hardware architectures. Recent efforts, such as \textsc{libCEED}~\cite{Ahmad2021} and \textsc{MFEM}, have focused on matrix-free and partially assembled high-order discretizations as core optimization strategies, particularly for GPU platforms, with \textsc{libCEED} serving as a tensor-product backend for \textsc{MFEM}'s high-order operator implementation. These efforts have shown notable application-level impact: \textsc{MFEM} formed part of the high-order finite-element infrastructure for the 2025 ACM Gordon Bell Prize-winning real-time tsunami forecasting application, which employed polynomial spaces up to fifth order~\cite{Henneking2025}. In contrast, high-order tensor-product optimization on CPUs has received comparatively less attention, with many implementations continuing to rely on general-purpose dense-kernel libraries, such as \textsc{LIBXSMM}~\cite{Heinecke2016}.

Graph coloring has long been used to eliminate write conflicts during shared-memory assembly and element updates, and communication overlap via thread-based progress has been extensively studied in MPI runtimes and applications~\cite{Jones1993,KRYSL2024117076,Cecka2011,Komatitsch2010}. While both graph coloring and progress threads are not novel concepts, our work integrates them into a unified SEM execution scheme for realistic wave-propagation workloads on ARM multicore CPUs, where CPU-side atomic updates are costly. Additionally, we employ a dedicated communication thread to ensure asynchronous progress under a hybrid MPI+OpenMP model~\cite{Rabenseifner2009,Gerald2011,Castillo2019}.

In comparison to prior high-order FEM/SEM optimization work, our contribution is twofold. First, we focus on ARM multicore CPUs with SME, a setting that has received less attention than GPU-oriented SEM optimizations. Second, rather than solely optimizing the tensor-product kernel, we demonstrate that SME shifts the performance-favorable operating point along the dispersion-based iso-accuracy $(h,p)$ frontier for wave-propagation SEM. Thus, our work is not merely a microarchitectural kernel optimization but an application- and architecture-aware co-design of discretization choices, kernel mapping, and hybrid parallel execution.

\section{Optimization}

\subsection{Aggregated SME Kernels for Batched Small Matrix Multiplications} \label{sec:SME_kernel}

\begin{figure*}[h]
     \centering
     \includegraphics[width=\linewidth]{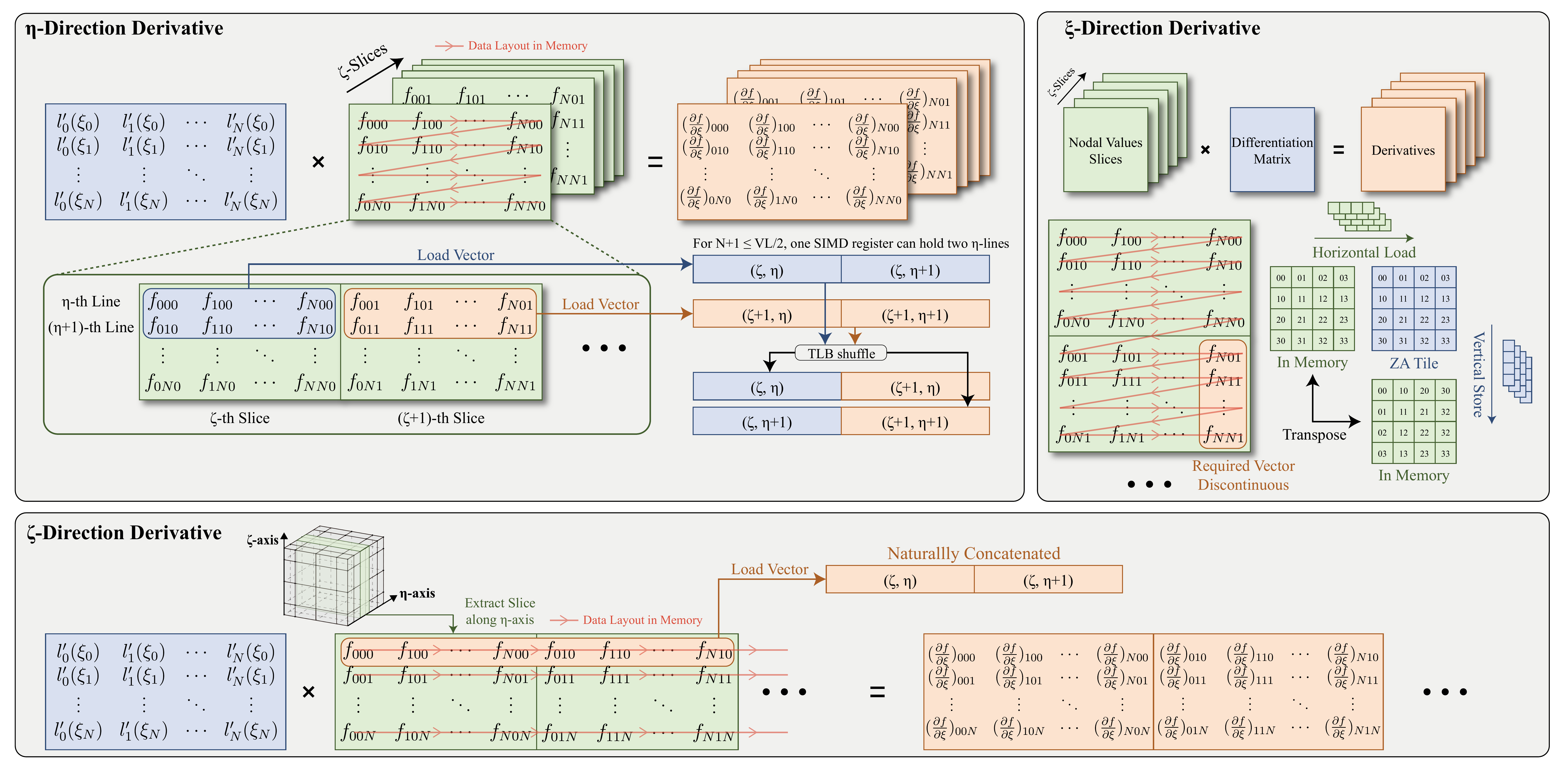}
     \caption{Aggregated SME-enabled Batched Matrix Multiplication}
     \label{fig:sme_aggr}
\end{figure*}

In this section, we address the mismatch between the SME ZA tile size and the dimensions of the batched small matrix multiplications in SEM kernels by combining software pipelining with SVE aggregation scheme. After the global degrees of freedom are gathered, each field component is stored in an element-local $(N{+}1)\times(N{+}1)\times(N{+}1)$ cube with the $\xi$-direction as the leading dimension (optionally with padding), and the one-dimensional differentiation operator is stored as an $(N{+}1)\times(N{+}1)$
coefficient matrix. We target polynomial orders from $p=4$, a common setting in existing SEM codes, up to $p=15$, which fully matches the ZA tile size for single-precision operands.

Among the three reference directions, the $\eta$-direction derivative most closely resembles a batched matrix-matrix multiplication, where each $\zeta$-slice of the element-local cube multiplies the one-dimensional coefficient matrix. SME exposes multiple ZA tiles that can be used for software pipelining to hide the latency of outer-product instructions. For large matrix multiplications, the four ZA tiles are typically grouped as a $2\times 2$ tile set for a single matrix pair, but this is unnecessary for batched small matrix multiplications whose dimensions are at most the vector length. Instead, we map different $\zeta$-slices to different ZA tiles and interleave outer products across the batch. Each tile processes a distinct slice, so all tiles remain active and outer-product latency is hidden by switching between slices, improving instruction throughput for the SEM workload.

We further increase ZA utilization within each $\zeta$-slice for low polynomial orders. Algebraically, this is equivalent to concatenating multiple $\zeta$-slices along the $\xi$-axis and applying the same coefficient matrix to the concatenated data. We realize
this logical concatenation on the fly using the vector-shuffling capabilities of the Scalable Vector Extension (SVE). For $N=4$-$7$, a single SIMD vector register can hold two $\eta$-lines. A vector load starting at one $\eta$-line brings both that line and the subsequent $\eta{+}1$ line. By combining this vector with the corresponding vector
from the next $\zeta$-slice and applying an SVE table-lookup shuffle, we construct operands for two consecutive outer-product iterations, so a single outer-product updates two $\zeta$-slices and effectively doubles ZA utilization.

The $\xi$-direction derivative has the same algebraic structure as the $\eta$-direction case: each $\zeta$-slice is multiplied by the same coefficient matrix. To improve SME utilization, we aggregate all $\zeta$-slices along the first dimension, forming a larger $((N+1)^2)\times(N+1)$ matrix in row-major order. However, this in-memory layout does not match the operand requirements of SME outer-product instructions, since each column is stored with non-unit stride. To avoid strided loads, we apply an in-register matrix transpose using the SME vertical and horizontal move instructions. This transformation reorganizes the data so that each column becomes contiguous in memory, enabling efficient SME outer-product evaluation.

For the $\zeta$-direction derivative, the operation can be viewed as a batched matrix multiplication between the coefficient matrix and a sequence of $\eta$-slices. We aggregate these $\eta$-slices along the second dimension, forming an $(N+1)\times((N+1)^2)$ matrix. In this arrangement, each row corresponds to a contiguous $\xi\eta$-plane in memory, so the resulting matrix multiplication matches the SME operand layout directly and can be executed without additional transpose or shuffle overhead.



We next derive a simple \emph{attainable-performance upper bound} for the SME-enabled kernels based on instruction-throughput considerations.

For the $\zeta$- and $\eta$-direction derivatives, evaluating all derivatives requires $N+1$ outer-product operations. Since the matrix dimensions are smaller than the SME tile size, each outer product utilizes only a fraction $\frac{N+1}{V_L}$ of the full tile capacity. In addition, $N+1$ outer-product updates are followed by $N+1$ store operations to write back the results. Under this simplified CPI-based view, the attainable floating-point throughput is upper-bounded by
\[
\frac{CPI_{OPA}}{CPI_{OPA}+CPI_{LSU}} \cdot \frac{N+1}{V_L}\, FLOPS_{SME}.
\]

For the $\xi$-direction derivative, the in-memory data layout does not match the operand layout required by SME outer-product instructions, so an additional transpose is needed. Using a rough throughput estimate, transposing an $(N+1)\times(N+1)$ matrix introduces another $N+1$ loads and $N+1$ stores through the ZA tiles. The corresponding attainable-performance upper bound is therefore reduced to
\[
\frac{CPI_{OPA}}{CPI_{OPA}+3\,CPI_{LSU}} \cdot \frac{N+1}{V_L}\, FLOPS_{SME}.
\]

Although simplified, these expressions make clear that before tile saturation, the maximum attainable SME throughput scales approximately linearly with $\frac{N+1}{V_L}$. This observation motivates the next subsection, where we examine how improved efficiency at higher polynomial orders can shift the favorable operating point on the $(h,p)$ design frontier.

\subsection{SME and the Performance-Favorable Operating Point on the $(h,p)$ Iso-Accuracy Frontier}

\begin{figure*}[h]
     \centering
     \includegraphics[width=\linewidth]{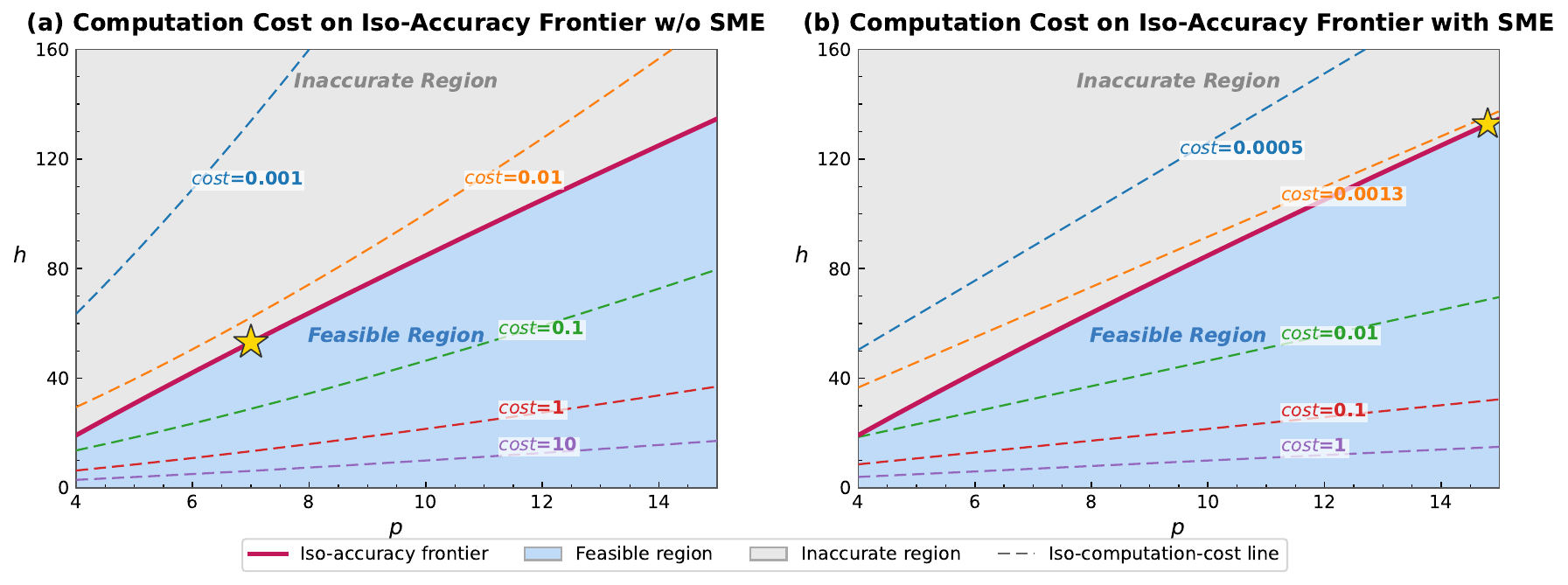}
     \caption{Shift of the Performance-Favorable Operating Point on the $(h,p)$ Iso-accuracy Frontier. Without SME, the dominant computational cost scales as $O\!\left(\frac{p^4}{h^3}\right)$, as shown in (a). With SME, improved utilization of tensor-product operators reduces the effective cost scaling to $O\!\left(\frac{p^3}{h^3}\right)$, shifting the favorable operating point toward higher polynomial orders, as shown in (b).}
     \label{fig:sme_iso_accuracy}
\end{figure*}

From \sect{SME_kernel}, our SME-enabled kernel maps the SEM batched small matrix multiplications to SME tiles with a utilization factor of $\frac{N+1}{V_L}$ and approaches the corresponding theoretical upper bound. Beyond this point, further improvement in SME efficiency cannot be achieved solely at the kernel level, and instead requires revisiting the application-level discretization choice.

For wave-propagation problems, the two dominant numerical error sources are temporal dispersion and spatial dispersion. The former arises from discrete time integration, while the latter is caused by insufficient spatial resolution to accurately represent high-frequency wave components. To reduce spatial dispersion, two standard refinement strategies are available: $h$-refinement, which decreases the element size, and $p$-refinement, which increases the polynomial order. Accordingly, the set of $(h, p)$ pairs that achieve the same spatial-dispersion accuracy defines an iso-accuracy frontier with respect to dispersion error.

For smooth solutions, SEM exhibits rapid convergence with increasing polynomial order, so higher $p$ can achieve the same spatial dispersion accuracy with fewer elements. Historically, however, the $O(N^4)$ cost of the tensor-product stiffness operator has made high-$p$ configurations less attractive in practice. SME changes this trade-off. Since SME efficiency increases approximately linearly with polynomial order before saturation at the vector length, the effective hardware cost scaling of the dominant batched matrix multiplication kernel is reduced from $O\!\left(\frac{N^4}{h^3}\right)$ to approximately $O\!\left(\frac{N^3}{h^3}\right)$ in this regime. As shown on Fig.~\ref{fig:sme_iso_accuracy}, the performance-favorable operating point along the iso-accuracy frontier shifts toward the higher-$p$ side.

Moving toward higher $p$ also brings an additional memory benefit. Because higher-order SEM can achieve the same spatial accuracy with fewer points per wavelength, the total number of global degrees of freedom can be reduced, leading to a smaller working set in memory. For ARM platforms such as LX2, where fast HBM capacity is limited, this directly translates into support for larger simulation domains or fewer required processors.

There is, however, an important counter-effect. The minimum nodal spacing, which governs the maximum allowable time step through the CFL condition, decreases as $O(N^{-2})$ for Gauss--Lobatto--Legendre nodes. In principle, this could require more time steps for a fixed physical simulation time and offset the gains from improved SME utilization. As we show in the experimental section using homogeneous equal-accuracy wave-propagation experiments however, for applications with high accuracy requirements such as reverse-time migration, the time-step restriction needed to control temporal dispersion is already more stringent than the CFL bound. In this regime, the use of higher polynomial order together with SME-enabled kernels becomes particularly attractive on ARM multicore platforms.

\subsection{Hybrid SEM Parallelization Scheme for Multicore CPUs}
\begin{figure}[h]
     \centering
     \includegraphics[width=\linewidth]{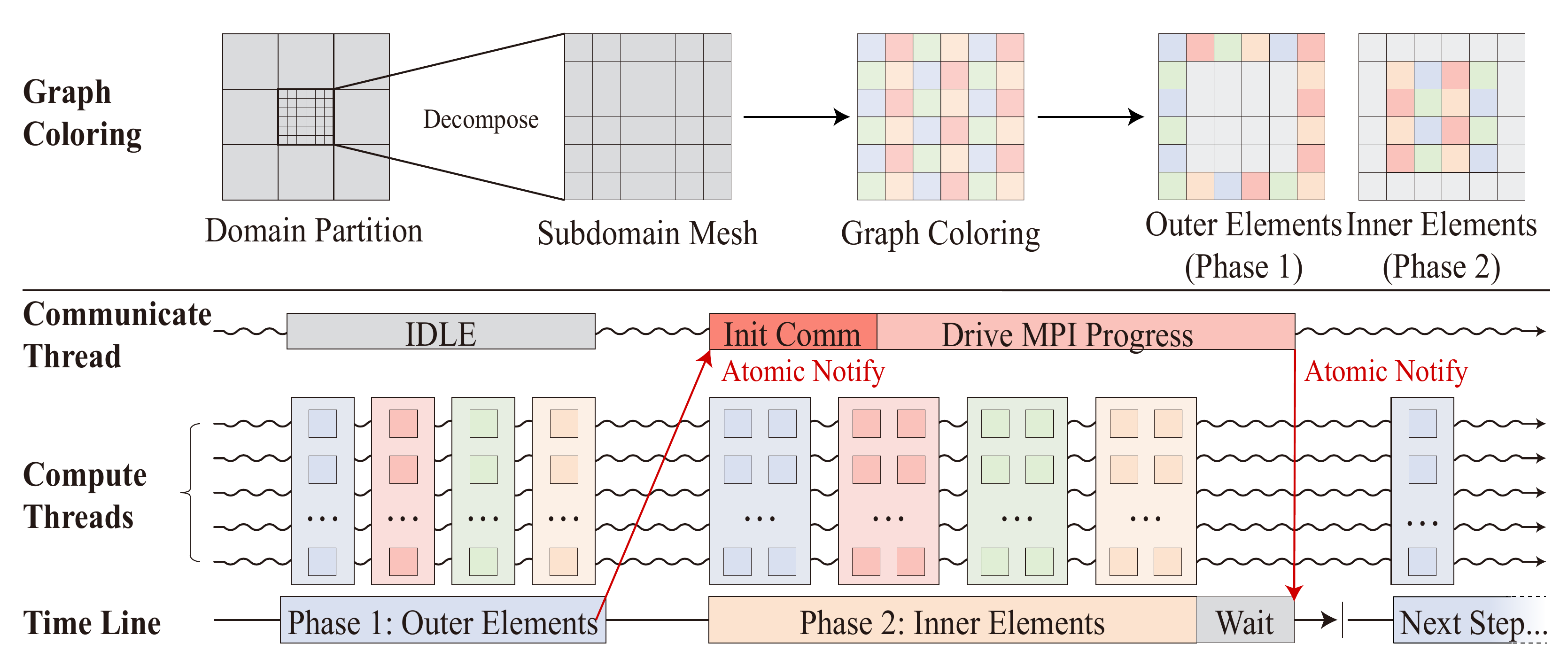}
     \caption{Hybrid Multiprocess/Multithread SEM Parallelization Scheme}
     \label{fig:hybrid_parallel}
\end{figure}

On modern multicore CPUs, growing core counts make MPI-only execution increasingly inefficient for SEM. In application frameworks such as \textsc{SPECFEM3D}, each process maintains substantial private data, including tomographic models, surface topography, and preprocessing metadata, so launching many MPI ranks creates significant pressure even on DDR memory.

This pressure is more severe at the NUMA and HBM levels. On LX2, each NUMA domain provides only 4~GB of HBM. With one MPI process per core, replicated boundary DoFs, communication buffers, and per-process metadata can consume a substantial fraction of this capacity, leaving little space for the PDE solve and, under huge-page mode, even risking allocation failure.




These considerations make hybrid parallelization essential for multicore CPUs. In our design, we map MPI processes to NUMA domains and employ multi-threaded parallelism within each NUMA. The whole workflow is shown in Fig.~\ref{fig:hybrid_parallel}.

On the NUMA-local, multi-threaded level, race conditions become a central issue, since adjacent elements share global nodes that must be updated consistently. In \textsc{SPECFEM3D}, OpenMP pragmas with atomic updates are used to remove these races. However, unlike GPGPUs, which often provide hardware floating-point atomics, CPUs implement atomics via load-linked/store-conditional (LL/SC) or compare-and-swap loops that repeatedly retry the update if another thread has modified the target location. Such atomic operations disrupt instruction pipelining and expose the full HBM access latency to the application, leading to poor scalability on multicore CPUs.

To avoid this overhead, we adopt a graph-coloring approach to eliminate race
conditions without atomics. \textsc{SPECFEM3D} uses a conformal hexahedral mesh, so elements can only be connected through shared vertices, edges, or faces. We exploit the existing global-to-local node mapping to construct an element-to-corner-vertex map, and then build an element adjacency graph via a sparse matrix-matrix multiplication on this incidence structure. A graph-coloring algorithm is applied to this adjacency graph, and elements are executed color by color. Within each color class, no two elements share nodes, so threads can update all nodal values without atomics or synchronization, removing shared-memory race conditions while preserving high pipeline and cache efficiency.

At the multi-process level, truly asynchronous neighbor exchanges become essential. However, the MPI progress model is often a limiting factor: nonblocking communication is only guaranteed to make progress when the application periodically invokes MPI routines such as \texttt{MPI\_test} or \texttt{MPI\_wait}. In practice, these calls may block the invoking thread or make only limited internal progress.

Given the abundant parallel resources on modern multicore CPUs, we dedicate one CPU core per MPI process to a progress thread responsible for handling all communication. The main thread initiates nonblocking neighbor exchanges and communicates with the progress thread via atomic variables. The progress thread repeatedly drives MPI progress on behalf of the process, while the main thread continues with computation. This design provides genuine asynchronous communication and enables full overlap between communication and computation for the SEM neighbor exchanges.





\subsection{Code-Specific Optimizations for SPECFEM3D} \label{sec:specfem3d_vector}

To further exploit the microarchitecture of the LX2, we refactor the core computing routines of \textsc{SPECFEM3D}.

First, for the kernel computing spatial derivatives, strains, and stresses, we fuse its multiple isolated spatial loops into a single grid-point traversal. This structural change eliminates intermediate gradient arrays, using scalars instead to avoid memory bandwidth bottlenecks. Furthermore, we move the element-regularity conditional completely outside the computational loops and specialize the kernel into two separate execution paths. By eliminating control-flow divergence within the hot loops, this transformation removes an obstacle to compiler vectorization.

Second, for the gather and scatter routines, we collapse the deeply nested 3D loops over the Gauss-Lobatto-Legendre points into a single 1D loop. This structural flattening enables higher SIMD utilization.
\section{Experimental Evaluation}

\subsection{Experimental Setup}

We use the LX2 platform described in \sect{kp920f} as our primary experimental system. Unless otherwise stated, all experiments are conducted on a single server equipped with two LX2 CPUs. We compile the code with the vendor-provided BiSheng~5.0.0 compiler and use Hybrid MPI~25.2.0. As the application baseline, we use the main branch of \textsc{SPECFEM3D} from GitHub at commit \texttt{7a1c764}. To ensure fair comparison across optimization stages and baselines, we fix the process/thread mapping, NUMA placement, and memory allocation policy throughout all experiments. 

For cross-platform comparison, we additionally use a server equipped with two Intel Xeon Gold~6248R processors (48 cores in total), 1.0~TiB DDR4 memory at 3200~MT/s, a 1.0~TiB NVMe SSD, and four NVIDIA A100-PCIE-40GB GPUs. We compile each implementation on different platforms with the highest stable optimization level together with architecture-specific code generation and fast-math enabled.

\subsection{Kernel-Level Evaluation of SME-Enabled Batched Small Matrix Multiplication}
\begin{figure*}[h]
     \centering
     \includegraphics[width=\linewidth]{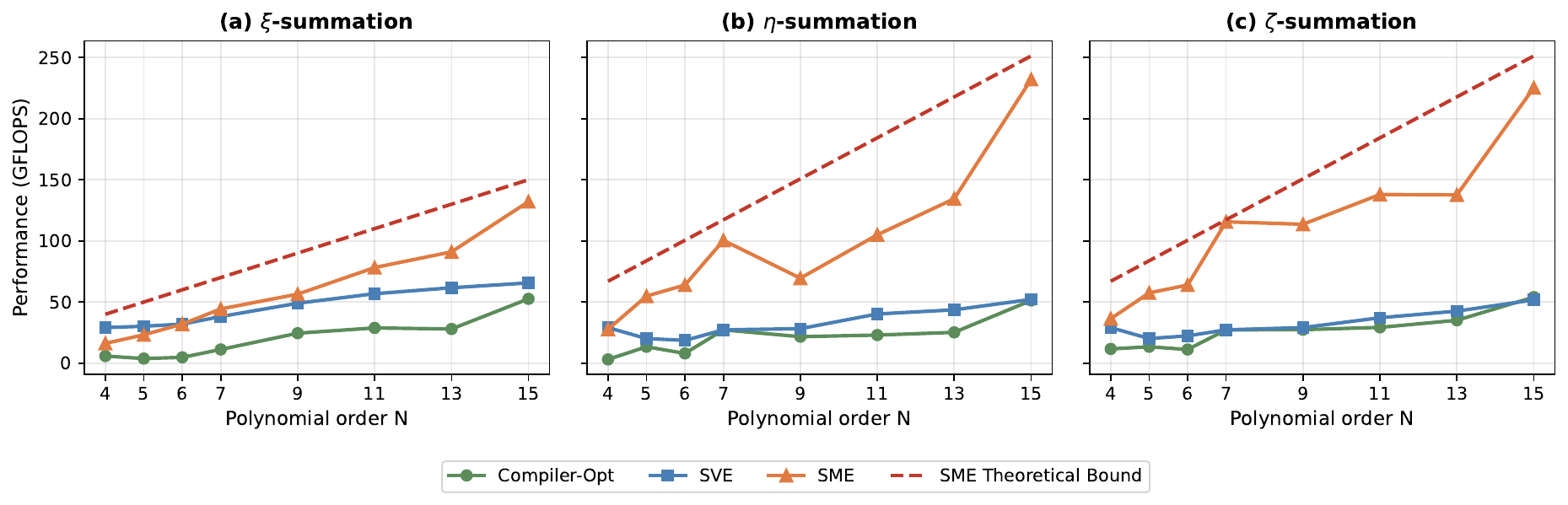}
     \caption{Batched Small Matrix Multiplication Performance }
     \label{fig:sme_summation}
\end{figure*}

\begin{figure*}[h]
     \centering
     \includegraphics[width=\linewidth]{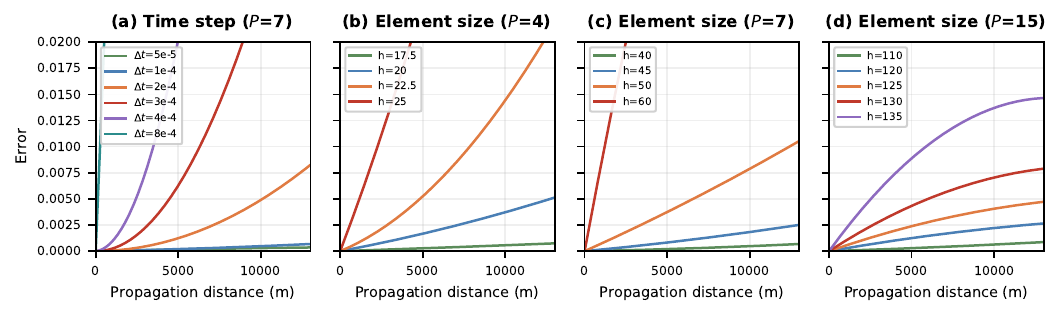}
     \caption{Wave Propagation Error on Different $(h,p)$ Settings}
     \label{fig:iso_accuracy}
\end{figure*}

In this subsection, we evaluate the proposed SME-enabled batched small matrix multiplication kernels. We use the effective single-core FLOP rate as the primary performance metric. As baselines, we include both the compiler-optimized implementation and a hand-written highly optimized SVE-intrinsic implementation\footnote{We additionally evaluated \textsc{LibXSMM} backend as a reference. However this generic backend was consistently below our specialized non-SME vectorized implementation due to lack of static shape information, so we report the latter as the strongest CPU baseline for assessing SME-specific gains.}. To assess how close the SME kernels are to the architectural limit, we also report the theoretical upper bound derived in \sect{SME_kernel}.

As shown in Fig.~\ref{fig:sme_summation}, the $\zeta$-direction kernel does not require transpose or explicit aggregation and therefore achieves the highest performance. For $p=7$ and $p=15$, where the matrix dimensions align well with the hardware, the achieved performance is close to the theoretical upper bound. In contrast, for the commonly used $p=4$ configuration, SME provides little advantage over SVE, which is consistent with the analysis in \sect{SME_kernel}. We also observe that, although the SVE kernel performance increases with $p$, it still reaches less than 60\% of the theoretical SVE compute capability. This is expected, since high SVE throughput requires sustained dual-FMA issue, whereas in small matrix multiplication the setup and store overheads remain significant.

For the $\eta$-direction kernel, the proposed aggregation scheme allows performance comparable to that of the $\zeta$-direction kernel. In these two directions, the advantage of SME over SIMD-based implementations is substantial: for $p=7$ and $p=15$, the SME kernels achieve up to $4\times$ speedup over the SIMD implementation. The $\xi$-direction kernel incurs additional transpose overhead and therefore performs worse than the other two directions. Nevertheless, it still outperforms the SIMD-based implementations when $p > 6$, and achieves approximately $2\times$ speedup at $p=15$.

\subsection{$(h,p)$ Dispersion-Based Iso-Accuracy Frontier Study} \label{sec:iso-ac_experiment}

\subsubsection{Equal-Accuracy Configuration Selection}
To quantify dispersion error and construct dispersion-based iso-accuracy configurations,  we consider acoustic wave propagation in a homogeneous medium and record the waveform at stations placed along one spatial direction. The simulation error is measured with respect to the analytical solution using the energy norm
\[
error(x) = \frac{\int \left(w(x,t)-w_{\mathrm{theory}}(x,t)\right)^2\,dt}
{\int w_{\mathrm{theory}}(x,t)^2\,dt}.
\]
The medium has density $1000~\mathrm{kg/m^3}$ and P-wave speed $1500~\mathrm{m/s}$. We use a Ricker source with dominant frequency $20~\mathrm{Hz}$, the acoustic solver in \textsc{SPECFEM3D}, and second-order Newmark time stepping.

Motivated by the accuracy requirements of reverse-time migration, we require that after $8~\mathrm{s}$ of propagation the waveform error remain below $1\%$. Since wave-propagation simulations are affected by both temporal and spatial dispersion, we first determine the maximum allowable time step by running on a sufficiently fine mesh such that spatial dispersion is negligible. As shown in Fig.~\ref{fig:iso_accuracy}, satisfying the target accuracy requires $\Delta t = 1\times10^{-4}~\mathrm{s}$.

We then determine the admissible element size for $p=4,7,15$ by testing different element sizes and using
\[
\Delta t = \min\left(\Delta t_{\mathrm{CFL}},\,1\times10^{-4}\right).
\]
In our high-accuracy setting, the temporal-dispersion constraint is more restrictive than the CFL limit. As shown in Fig.~\ref{fig:iso_accuracy}, the maximum admissible element size is $20~\mathrm{m}$ for $p=4$, $50~\mathrm{m}$ for $p=7$, and $130~\mathrm{m}$ for $p=15$.

\subsubsection{Computational and Memory Cost on the Iso-Accuracy Frontier}
Using the equal-accuracy $(h,p)$ configurations derived above, we evaluate isotropic elastic wave propagation and compare two pairs of discretizations: $p=4$ versus $p=7$, and $p=7$ versus $p=15$, with our optimized \textsc{SPECFEM3D}. The goal is to quantify the additional application-level benefit that becomes available once SME-aware execution makes higher-order discretizations practically attractive. In each case, the computational domain is chosen to saturate the available HBM resources under the lower-order configuration, and the element sizes for the higher-order configurations are scaled according to the equal-accuracy ratios determined above.

Because the temporal-dispersion constraint is stricter than the CFL stability limit in this regime, we use the same time step for all configurations and run the \textsc{SPECFEM3D} solver for 15{,}000 steps, recording the total simulation time.

\begin{figure}[h]
     \centering
     \includegraphics[width=\linewidth]{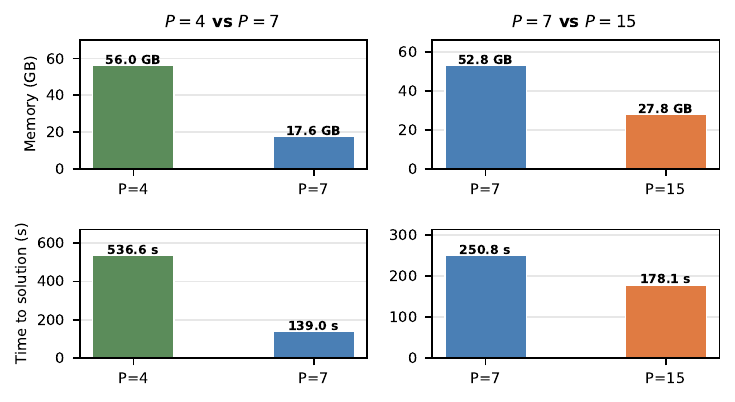}
     \caption{SME-enabled Performance Comparison on Iso-Accuracy Frontier}
     \label{fig:performance_iso_accuracy}
\end{figure}

As shown in Fig.~\ref{fig:performance_iso_accuracy}, switching from $p=4$ to $p=7$ yields a $3.18\times$ reduction in memory usage and a $3.85\times$ reduction in time-to-solution. The memory reduction arises because higher polynomial order can achieve the same accuracy with fewer points per wavelength, thereby reducing the total number of global degrees of freedom. This is particularly important for applications such as reverse-time migration, where wavefield history must be stored, and for platforms such as ARM multicore CPUs with limited HBM capacity. Under this higher-order setting, workloads that cannot fit within a single server node at low polynomial order can be executed within one node.

The trend continues when moving from $p=7$ to $p=15$, yielding a further $1.8\times$ reduction in memory usage and a $1.4\times$ reduction in time-to-solution. These results support the use of higher polynomial order together with SME-enabled kernels for wave-propagation simulations on ARM multicore platforms.

\subsection{End-to-End Performance of Optimized \textsc{SPECFEM3D}}

\begin{figure}[h]
     \centering
     \includegraphics[width=\linewidth]{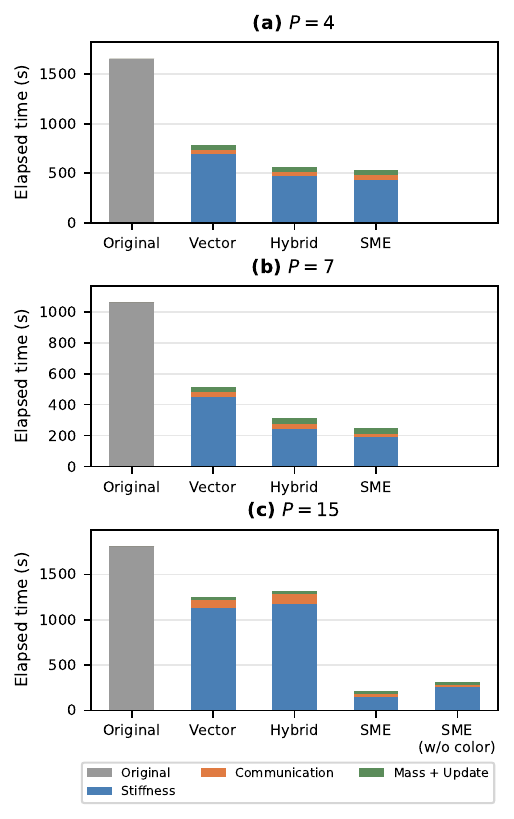}
     \caption{Performance Breakdown}
     \label{fig:performance_breakdown}
\end{figure}

\subsubsection{Breakdown of Optimization Contributions}

We evaluate the contribution of each optimization to reducing time-to-solution under three polynomial-order settings: $p=4,7,15$. The earlier stages quantify the benefit of general CPU-side optimization and hybrid execution at fixed discretization, whereas the final stage isolates the incremental contribution of SME relative to previous stages. For each setting, we choose a mesh that saturates the available HBM capacity on one server node and run the simulation for 15{,}000 steps.

The vectorization optimization in \sect{specfem3d_vector} improves performance by $2.1\times$, $2.3\times$, and $1.4\times$ over the original \textsc{SPECFEM3D} for $p=4$, $p=7$, and $p=15$, respectively, and thereby establishes a stronger non-SME baseline for the subsequent stages. Graph coloring further improves performance by $1.40\times$ and $1.47\times$ for $p=4$ and $p=7$, but yields only $0.94\times$ for $p=15$. The degradation at $p=15$ can be explained by two factors. First, when the batched small matrix multiplication becomes sufficiently expensive, the cost of serialized atomic updates accounts for a smaller fraction of the total runtime. Second, executing the stiffness operator color by color reduces the effective working set of each OpenMP region and weakens the locality benefits introduced by DoF renumbering.

The final stage adds SME on top of the optimized non-SME implementation. For the stiffness operator alone, SME yields speedups of $1.06\times$, $1.27\times$, and $7.69\times$ for $p=4$, $p=7$, and $p=15$, respectively; these translate into end-to-end simulation speedups of $1.05\times$, $1.26\times$, and $5.92\times$. The improvements at $p=4$ and $p=7$ are consistent with the kernel-level results. For $p=15$, the observed application-level speedup is larger than the kernel-level SME gain alone would suggest. The reason is that the baseline implementation uses a generic runtime-parameterized batched matrix multiplication path for polynomial orders higher than 10, whereas our optimized implementation combines SME-specific acceleration with static shape specialization. Accordingly, the improvement at $p=15$ reflects the joint benefit of architectural support and specialization to SEM operator structure.

The communication thread provides a mechanism to overlap communication with computation. In practice, however, its impact is limited. Domain decompositions produced by METIS or Scotch typically introduce a workload imbalance of about 5--10\%, and under the hybrid parallelization scheme each process exchanges less than 1~MB of halo data compared to a working set of several gigabytes. Consequently, the remaining asynchronous wait time is dominated by load imbalance rather than communication overhead, and the communication-thread optimization contributes little to end-to-end performance.

Finally, to confirm the effectiveness of graph coloring at $p=15$, we conduct an ablation study on the final optimized version by removing graph coloring and reverting to OpenMP atomic updates. This increases the simulation time by $1.16\times$, demonstrating that graph coloring remains beneficial once the kernel is optimized.

\subsubsection{Scaling Experiments}
\begin{figure}[h]
     \centering
     \includegraphics[width=\linewidth]{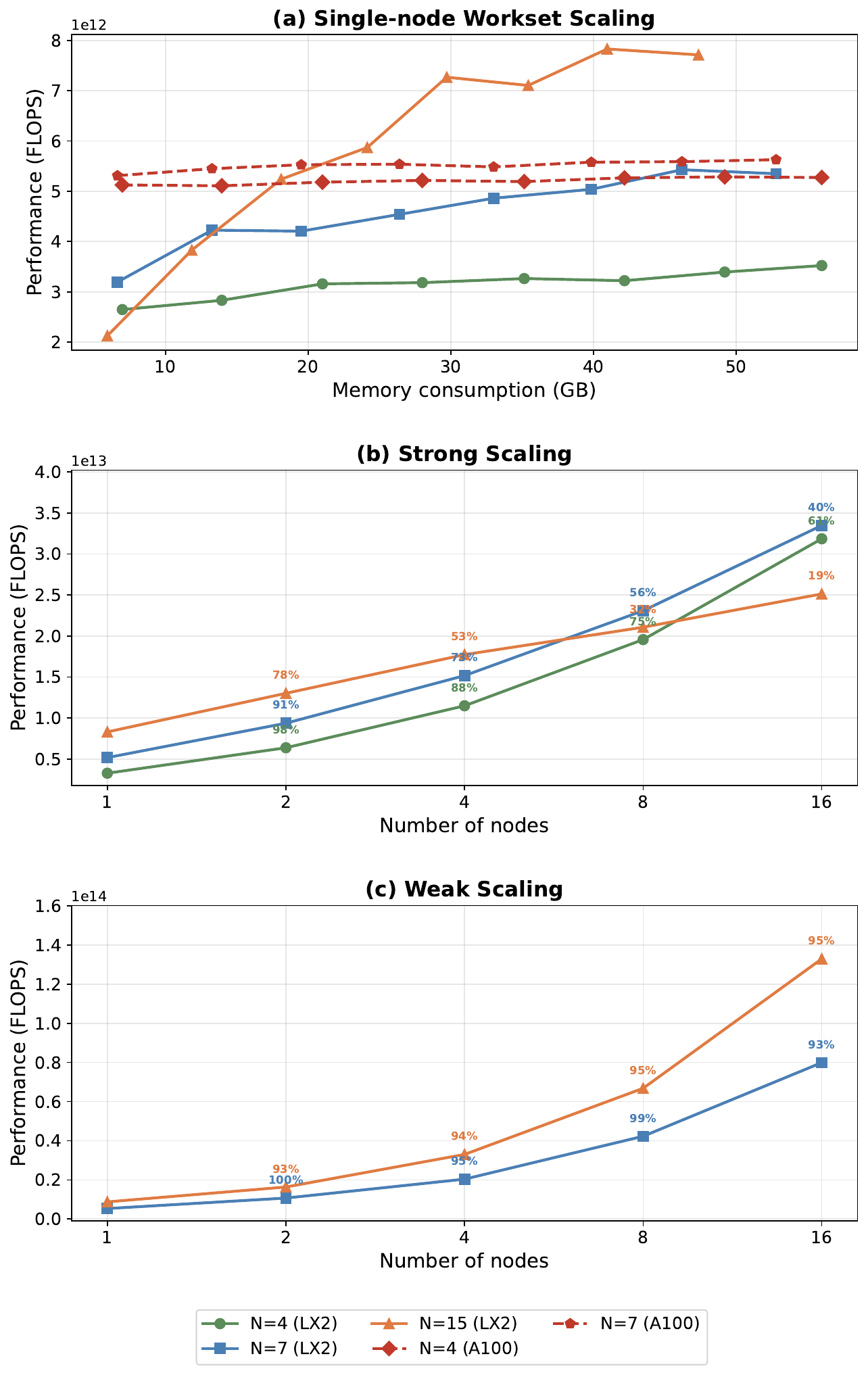}
     \caption{Intra/Inter-node Scaling Experiments}
     \label{fig:scaling}
\end{figure}
For the intra-node scaling experiments, we additionally include a cross-platform comparison against two NVIDIA A100 GPUs using the same mesh configuration. To present results for different polynomial orders in a single figure, we use the memory footprint on a LX2 server node as the horizontal axis.

As shown in Fig.~\ref{fig:scaling}, higher polynomial order increasingly favors the ARM platform because SME utilization improves with $p$. For $p=4$, the A100 is approximately twice as fast, reflecting the low SME utilization in this regime. At $p=7$, LX2 becomes competitive with the A100. At $p=15$, the A100 implementation can no longer be compiled because of excessive shared-memory demand, whereas SME reaches near-full utilization and delivers approximately $1.6\times$ higher FLOP rate than at lower order. Our cross-platform results show that SME materially improves the competitiveness of ARM multicore CPUs for tensor-product SEM, reducing the gap to accelerator-oriented platforms and changing the practical role of CPUs from baseline hosts to viable high-performance execution targets. Graph coloring reduces the number of elements processed within each parallel region, making performance more sensitive to working-set size; if the local workset becomes too small, hardware resources cannot be fully saturated. From this perspective, the higher-order cases in \sect{iso-ac_experiment} have not yet reached their full performance potential, and the gain from larger $p$ would likely increase further for larger worksets.

For the inter-node weak-scaling study, we report only $p=7$ and $p=15$, since the $p=4$ case requires too many elements and exceeds the capability of our mesher. The weak-scaling results are close to ideal. As discussed for the communication-thread optimization, the halo region is small under the hybrid execution model: each process exchanges less than 1~MB of data compared with a working set of several gigabytes. As a result, communication overhead is minor, and the observed deviation from ideal weak scaling is dominated instead by system noise and by the 5--10\% load imbalance introduced by the METIS/Scotch partitioning.

For strong scaling, the problem size is fixed while the number of resources is increased, so the work per process decreases and parallel overhead becomes more visible. Under this regime, $p=15$ exhibits the weakest strong scaling because our optimizations are sensitive to local element count and working-set size, whereas $p=4$ shows the strongest strong-scaling behavior. This is consistent with the standard observation that strong scaling degrades once the local workload becomes too small to keep the hardware fully utilized.

\section{Conclusion}
In this work, we optimized \textsc{SPECFEM3D} for ARM multicore CPUs with SME and showed that the benefit of SME extends beyond faster tensor-product kernels. By combining an SME-aware batched small-matrix implementation with a memory-aware hybrid MPI+OpenMP execution scheme, we substantially improved full-application performance on LX2 while enabling realistic workloads under limited HBM capacity. More importantly, through a dispersion-based iso-accuracy study, we showed that SME changes the practical cost trend across polynomial orders and makes higher-$p$ SEM discretizations significantly more attractive on modern ARM multicore CPUs. These results suggest that matrix-oriented CPU extensions can affect not only kernel mapping, but also application-level discretization choices in high-order wave-propagation codes.

\bibliographystyle{IEEEtran}
\bibliography{reference}

\end{document}